%% file: samplepaper.tex
\documentclass[runningheads]{llncs}
\usepackage[T1]{fontenc}
\usepackage{graphicx}
\usepackage[most]{tcolorbox}
\usepackage{url}
\usepackage{hyperref}

\begin{document}
\title{A Multi-Agent Framework for Democratizing XR Content Creation in K-12 Classrooms}
%
%
\author{%
Yuan Chang\inst{1}\textsuperscript{*} \and
Zhu Li\inst{2}\textsuperscript{*} \and
Jiaming Qu\inst{3}\textsuperscript{*}
}

\authorrunning{Y. Chang et al.}
%
\institute{Meta, Burlingame, CA, USA \email{yuanchang96@meta.com} \and
Meta, Burlingame, CA, USA \email{zhuli@meta.com} \and
Amazon, Seattle, WA, USA \email{qjiaming@amazon.com}}

\maketitle              
\begingroup
\setcounter{footnote}{0}
\footnotetext[1]{Authors are listed alphabetically. This work was conducted in a personal capacity and does not reflect the views of the authors' employers.}
\endgroup

\input{tex/abstract}

\input{tex/introduction}
\input{tex/related_work}
\input{tex/methods}
\input{tex/case_study}

\input{tex/discussion}

\bibliographystyle{splncs04}
\bibliography{reference}

\end{document}

%% file: tex/abstract.tex
\begin{abstract}

Generative AI (GenAI) combined with Extended Reality (XR) offers potential for K-12 education, yet classroom adoption remains limited by the high technical barrier of XR content authoring. Moreover, the probabilistic nature of GenAI introduces risks of hallucination that may cause severe consequences in K-12 education settings. In this work, we present a multi-agent XR authoring framework. Our prototype system coordinates four specialized agents: a Pedagogical Agent outlining grade-appropriate content specifications with learning objectives; an Execution Agent assembling 3D assets and XR contents; a Safeguard Agent validating generated content against five safety criteria; and a Tutor Agent embedding educational notes and quiz questions within the scene. Our teacher-facing system combines pedagogical intent, safety validation, and educational enrichment. It does not require technical expertise and targets commodity devices.

\keywords{Generative AI  \and Extended Reality \and K-12 Education.}
\end{abstract}

%% file: tex/introduction.tex
\section{Introduction}

Extended Reality (XR), which includes virtual reality, augmented reality, and mixed reality, is a technology that blends digital elements with the physical world using special devices. Prior studies have shown that XR technologies have been widely used in classrooms~\cite{khukalenko2022teachers,meccawy2022creating}. Compared to conventional education materials, immersive environments can support engagement, spatial understanding, and experiential learning in ways that are difficult to achieve through pure textbooks or videos alone.

Despite this potential, XR remains difficult to integrate into everyday classroom practice due to several challenges~\cite{rossi2024impact,linares2024breaking}. First, effective use of AI tools often requires specialized expertise, such as prompt engineering or model configuration, which many educators lack. Second, AI-generated content may be unsafe, inaccurate, or ethically problematic, including the presence of violence, bias, or hallucinated facts. These risks are particularly harmful in specific settings like K-12 education. Meanwhile, the development of immersive learning experiences typically requires substantial technical skills and access to specialized hardware, further limiting classroom uptake. As a result, teachers struggle not with a lack of available content, but with a lack of AI-assisted tooling with control and usability.

In this paper, we focus on XR content authoring in K-12 education. Using Large Language Models (LLMs), we developed a multi-agent system involving pedagogical interpretation, asset generation, content review, and instructional enrichment. This system enables teachers to describe a desired learning experience in plain language and obtain an XR-based educational scene with instructional support. We designed our system as a teacher-facing, browser-based workflow that runs on commodity devices and reduces dependence on technical XR authoring expertise.

Our work makes three main contributions. First, we present a K-12-oriented framework for AI-assisted XR authoring in which pedagogical intent, safety, and instructional usability must be addressed jointly. Second, we introduce a multi-agent architecture that implements this framework in a human-in-the-loop pipeline. Third, we release a working system that demonstrates the end-to-end feasibility of generating classroom-oriented XR learning artifacts.

We release our codebase at: \url{https://github.com/cruisekkk/K12_XR}.

%% file: tex/related_work.tex
\section{Related Work}

\textbf{The Evolution of XR:} XR environments and applications have traditionally been created through expert workflows: designers create 3D assets, assemble scenes in engines, and implement interaction logic through code, followed by repeated build-deploy-test cycles on target hardware~\cite{doolani2020review}. To reduce the technical barriers for XR development using specialized frameworks and platforms, recent studies have explored alternative approaches (e.g., web-based authoring tools~\cite{krings2022fader} and AI-powered XR authoring~\cite{lee2025imaginatear}) that prioritize end-user needs and simplified controls. Another approach uses pattern-based or template-based authoring, where reusable elements keep experiences executable during creation and support incremental customization~\cite{rau2022pattern,horst2022authoring}.

\textbf{LLMs for XR:} There is growing interest in using LLMs as co-creators in XR authoring to translate user intent into interactive environments. For instance, Torre et al. developed LLMR, a framework enabling the real-time creation and modification of mixed-reality scenes in the Unity engine via text prompting. This system combines task planning, self-debugging, and memory mechanisms, and the authors reported positive usability~\cite{de2024llmr}. Similarly, Giunchi et al. designed DreamCodeVR. This system assists users without any prior programming knowledge to craft basic object behavior in VR environments by translating spoken language into code within an active application~\cite{giunchi2024dreamcodevr}. Beyond code generation and content creation, prior studies have integrated LLMs into XR scenes as conversational assistants that provide real-time feedback~\cite{elfleet2024investigating,w2025latency}.

\textbf{LLMs in K-12 Education:} K-12 education settings include specific constraints (e.g., minors, curriculum standards, and assessment integrity) that pose unique challenges for how LLMs can be responsibly introduced. For instance, studies have shown that teachers already use LLMs for preparation work (e.g., quizzes, lesson plans, slide decks), but largely as short, single-mode interactions rather than sustained collaboration~\cite{keppler2025making}. Co-design with K-12 project-based learning educators likewise suggests LLM tools should automate routine logistics while preserving teacher autonomy and highlighting classroom constraints and ethical concerns~\cite{ravi2025co}. On the learner side, studies have shown that LLM-generated explanations can match expert-authored supports for middle-school math. Additionally, persona-based systems have potential for dialogic engagement and empathy in history education, while also revealing limitations regarding reliability and classroom fit~\cite{worden2025scaling,kim2025histochat}.

%% file: tex/methods.tex
\section{Methods}

\subsection{Design Process}
Our system design was based on the observation that XR content authoring for education involves four key challenges.

\begin{enumerate}
    \item \textbf{Reduce authoring burden for non-technical teachers.} Teachers typically formulate classroom needs in pedagogical language rather than in the technical language required by 3D generation systems. A useful authoring system must therefore bridge this gap without expecting teachers to become prompt engineers or XR developers.
    \item \textbf{Preserve pedagogical intent and teacher agency.} In classroom settings, teachers remain responsible for choosing what students should learn and how materials should support that learning. AI assistance should therefore not replace pedagogical decision-making but help translate teacher intent into usable artifacts.
    \item \textbf{Treat classroom safety as an explicit design concern.} In K-12 settings, content suitability cannot be assumed from visual plausibility alone. Generated artifacts must be reviewed for age appropriateness, harmful content, factual consistency, bias, and educational relevance.
    \item \textbf{Produce teachable artifacts, not just 3D assets.} A 3D model alone is often insufficient for classroom use. Teachers need materials that include explanations, vocabulary, annotations, and opportunities for formative assessment. Educational XR authoring should therefore return instructional scaffolds alongside the generated visual content.
\end{enumerate}

These requirements motivated our multi-agent design. Rather than using one single model, we decompose XR authoring into specialized stages that reflect the workflow of specifying, generating, reviewing, and enriching educational content.

\subsection{System Architecture Overview}

\begin{figure*}[htbp!]
\centering
\includegraphics[width=.6\textwidth]{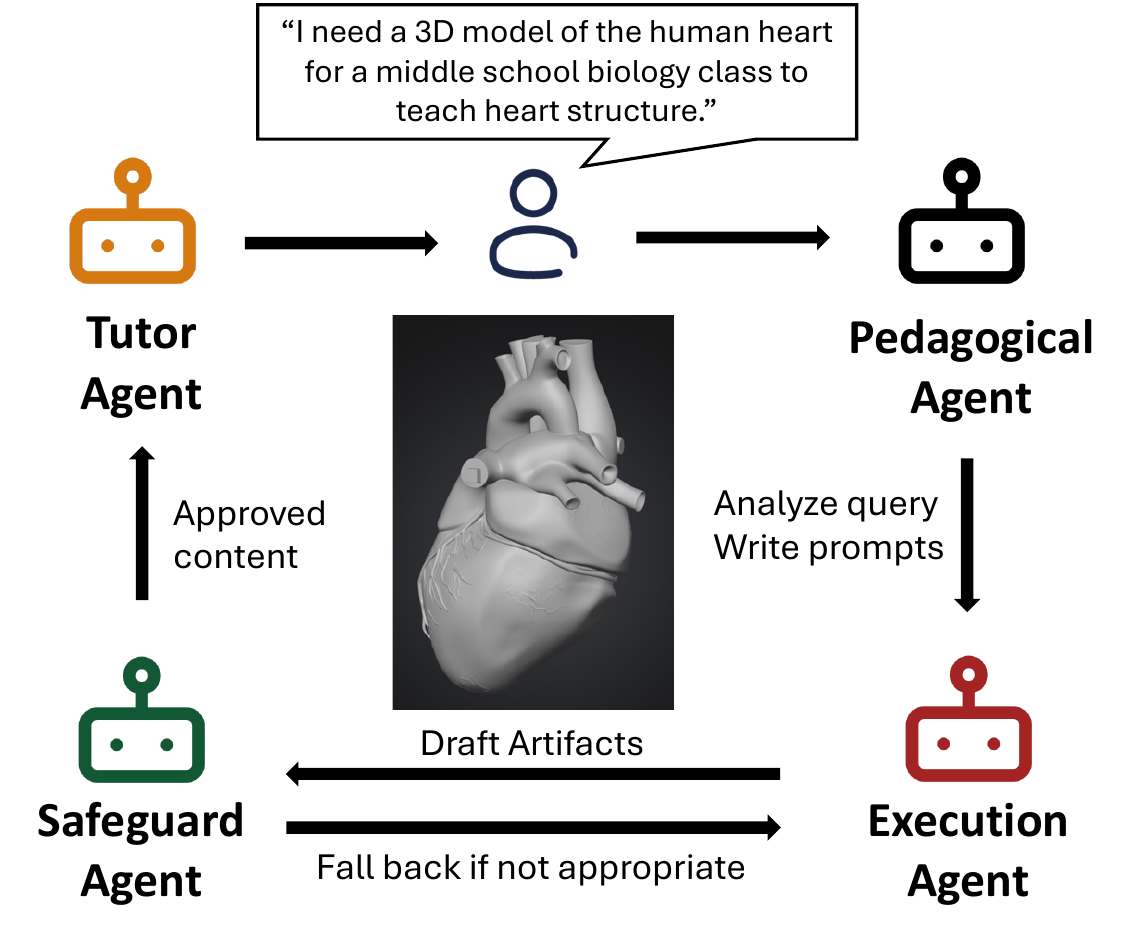}
\caption{System architecture overview.}
\label{system_overview}
\end{figure*}

As shown in Figure~\ref{system_overview}, our system is organized as a sequential workflow that transforms a teacher's natural-language request into an XR-ready educational scene. First, the system interprets the teacher's request pedagogically, producing a structured content brief with learning goals and grade-appropriate detail. Second, it converts that specification into a 3D asset through an external generation pipeline. Third, it reviews the resulting content against defined K-12 safety criteria, with the option to revise and retry generation if necessary. Finally, it enriches the approved scene with annotations and supporting instructional content.

Our system has two core parts: (1) a front-end web-based interface where the user can interact with the system to configure the prompt and get the generated XR content, and (2) a backend system operationalizing the four agents and detailed workflows. We implemented the system using the following technology stack: Next.js, React, and Zustand for the front-end interface; Google model-viewer for rendering 3D/XR contents; and FastAPI and Uvicorn for the back-end system. For the agents, we provided endpoints to use Claude or OpenAI series LLMs, depending on the user's preference.

\subsection{Implementation of Agents}
In this section, we detail the implementation of the four agents.

\textbf{Pedagogical Agent:} The Pedagogical Agent addresses a key mismatch in XR authoring: teachers express goals in instructional terms, whereas 3D generation systems require visual and structural specificity. This agent reformulates the teacher's request into a structured content specification. This design offloads the challenge of crafting complex prompts from teachers. More specifically, we prompt an LLM to act as a K-12 curriculum expert to identify the core educational concept, determine the right level of detail for the grade level, and produce a refined prompt optimized for 3D generation using the following instructions:

\begin{tcolorbox}[
    colback=black!5,
    colframe=black!40,
    arc=2mm,
    boxrule=0.8pt,
    left=2mm,
    right=2mm,
    top=2mm,
    bottom=2mm,
    title=\textbf{System Prompt for the Pedagogical Agent}
]
\small
You are a K-12 Pedagogical Expert Agent. Your role is to transform simple teacher prompts into detailed, curriculum-aligned prompts for 3D content generation.

Given a teacher's natural language request, you must:

1. Identify the core educational concept

2. Determine the appropriate detail level for the grade level

3. Generate a refined, detailed prompt optimized for 3D model generation

Your refined prompt should specify:

- Scientific/factual accuracy requirements

- Key visual features and structures that must be present

- Appropriate complexity for the target grade level

- Real-time rendering optimization notes

- Educational labeling requirements
\end{tcolorbox}

\textbf{Execution Agent.} The Execution Agent is responsible for converting the pedagogical specification into a 3D asset. At this stage, we leverage the Meshy API\footnote{https://www.meshy.ai/} to produce a 3D model in the graphics library transmission file format using the prompt generated by the Pedagogical Agent. Separating execution from pedagogical interpretation allows the system to keep generation concerns distinct from instructional reasoning.

\textbf{Safeguard Agent.} We treat safety review as an explicit process stage rather than an implicit expectation of the generator. The Safeguard Agent evaluates content across five defined K-12 dimensions: age appropriateness, factual accuracy, absence of violent or disturbing imagery, absence of racial, gender, or cultural bias, and educational alignment. To this end, we prompt an LLM to act as a reviewer to carefully inspect both the textual specification and a rendered image of the generated content. If the output fails the review, the pipeline re-enters the generation stage, using safeguard feedback to guide the next attempt.

\begin{tcolorbox}[
    colback=green!5,
    colframe=green!40,
    arc=2mm,
    boxrule=0.8pt,
    left=2mm,
    right=2mm,
    top=2mm,
    bottom=2mm,
    title=\textbf{System Prompt for the Pedagogical Agent}
]
\small
You are a K-12 Content Safety Agent. Your role is to evaluate generated educational content for appropriateness in K-12 classroom settings.

You must check for:

1. Age-appropriateness: Content must be suitable for the specified grade level

2. Accuracy: Scientific/factual content should be accurate and not misleading

3. Safety: No violent, sexual, or disturbing imagery

4. Bias: No racial, gender, cultural, or other biases

5. Educational value: Content should support learning objectives

Be strict about safety but reasonable about educational content. Medical/anatomical models are acceptable when scientifically accurate and age-appropriate.
\end{tcolorbox}

\textbf{Tutor Agent.} The Tutor Agent transforms a generated asset into a more teachable learning artifact. It augments the scene with educational annotations, lesson overviews, vocabulary definitions, quiz questions, and related learning materials. We used Tavily\footnote{https://www.tavily.com/} as the backbone of the agent. This agent first retrieves supporting information through a web search and then uses an LLM to synthesize classroom-oriented scaffolds grounded in that context, such as an overview, glossary, quiz questions, and reading materials.

\begin{tcolorbox}[
    colback=orange!5,
    colframe=orange!40,
    arc=2mm,
    boxrule=0.8pt,
    left=2mm,
    right=2mm,
    top=2mm,
    bottom=2mm,
    title=\textbf{System Prompt for the Pedagogical Agent}
]
\small
You are a K-12 Educational Tutor Agent. Your role is to create rich educational content that accompanies 3D models in an XR learning environment.

Given information about a 3D model and its subject matter, you must generate:

1. Educational annotations that can be placed on/near the 3D model

2. A structured lesson overview

3. Interactive quiz questions

4. Key vocabulary terms with definitions
\end{tcolorbox}

%% file: tex/case_study.tex
\section{Illustrative Walkthrough}

\begin{figure*}[ht!]
\centering
\includegraphics[width=\textwidth]{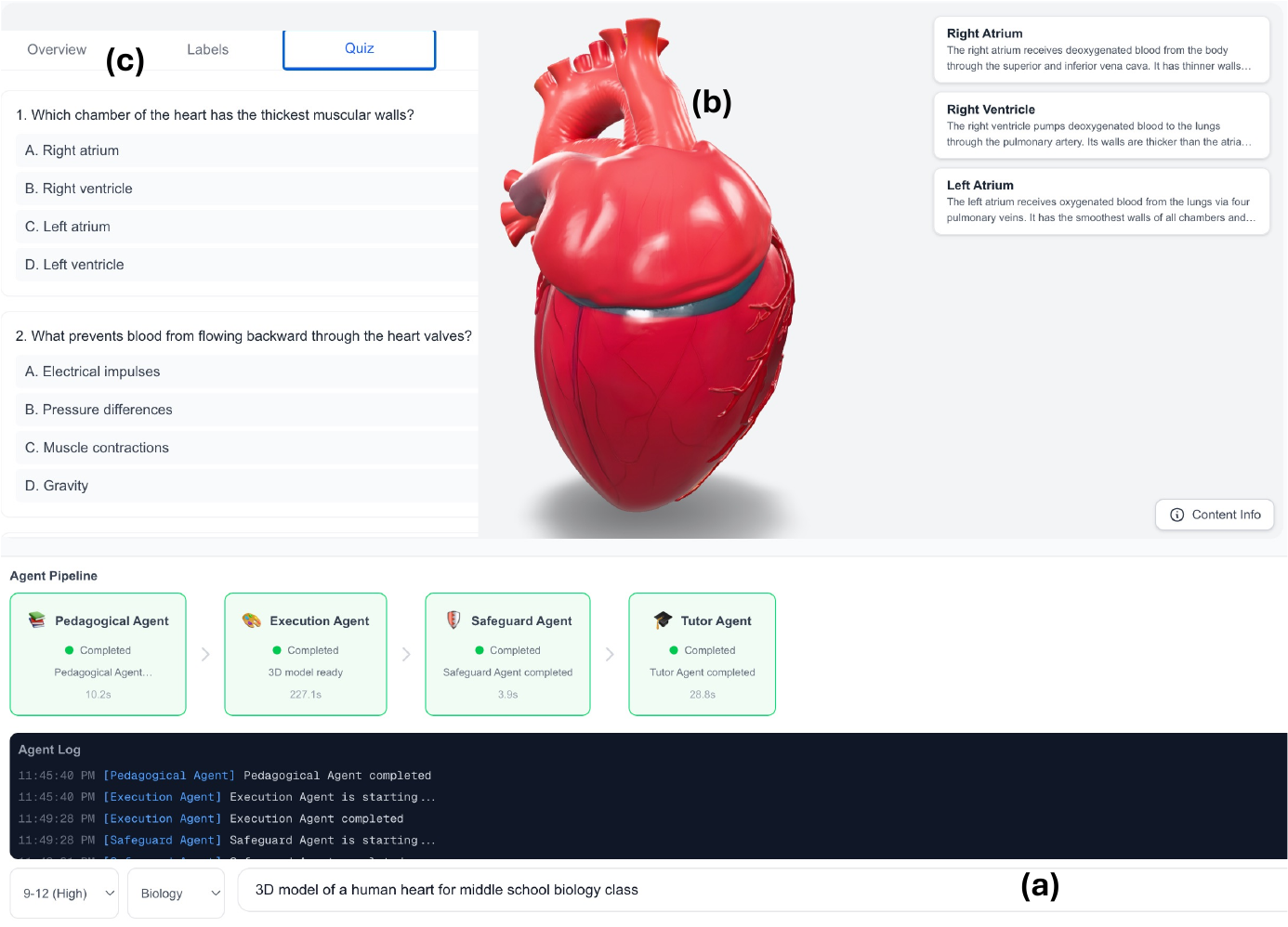}
\caption{System interface. (a) The user can configure education level, subject, and topic of interest, which will be passed to the pedagogical agent. (b) The generated 3D model generated and checked by the execution and safeguard agents. (c) Teaching materials (e.g., introduction and quiz questions) generated by the tutor agent.}
\label{example}
\end{figure*}

To illustrate the workflow, consider a middle school biology teacher requesting ``a 3D model of a human heart for a middle school biology class.'' As shown in Figure~\ref{example}, the system first interprets this prompt pedagogically. Rather than sending the original request directly to a generator, the Pedagogical Agent reformulates it into a more structured specification that includes grade-appropriate learning goals and key anatomical features relevant to the lesson. The Execution Agent then produces a browser-viewable heart model. Next, the Safeguard Agent reviews the generated content for classroom suitability, examining whether the model is appropriate for the intended grade level, visually non-disturbing, factually consistent, and educationally useful. Once approved, the Tutor Agent enriches the scene with annotations for major structures, vocabulary items, and quiz questions that can support guided exploration or formative assessment. A full video walkthrough of this example is available \href{https://jiamingqu.com/HCII26_demo}{\textcolor{blue}{\underline{here}}}.

%% file: tex/discussion.tex
\section{Discussion}

\textbf{Summary:} In this paper, we present a multi-agent framework for making XR content creation accessible in K-12 classrooms. By coordinating four specialized AI agents---Pedagogical, Execution, Safeguard, and Tutor---in a sequential pipeline, the system enables teachers to generate safe, curriculum-aligned, interactive 3D educational scenes from simple natural language prompts, without requiring expertise in 3D modeling or prompt engineering. Unlike existing XR authoring tools or general-purpose LLM interfaces, the system combines pedagogical intent capture, defined multi-criteria safety validation, and automated educational enrichment in a single, teacher-facing pipeline that runs on commodity hardware. By offloading technical complexity to agents while retaining pedagogical oversight, we offer a scalable and equitable approach to integrating immersive technology into everyday K-12 instruction.

\textbf{Design Implications:} Our multi-agent system has the potential to translate natural language pedagogical intent into safe, deployable XR content in K-12 classrooms. It provides the following design implications for future systems. First, we integrated responsible AI practices in high-stakes educational contexts. The separation of concerns across specialized agents suggests a general architectural pattern for AI-assisted content creation workflows that require both domain expertise and safety guarantees. Second, the decision to target commodity hardware (e.g., browsers that are accessible on any tablet or PC) improves equity in immersive learning experiences. Compared to specialized hardware like headsets, this approach provides opportunities to use XR contents in classrooms in under-resourced areas. Finally, the human-in-the-loop design, in which teachers specify intent, review the Pedagogical Agent's interpretation, and observe Safeguard Agent decisions, preserves teacher agency and positions AI as a tool that augments rather than replaces educator judgment.

\textbf{Limitations and Future Work:} The current system is a proof-of-concept prototype with several important limitations. First, the 3D generation pipeline relies on the Meshy API, which can take one to five minutes per model and incurs per-generation costs that may be expensive at scale. Second, the Safeguard Agent evaluates content based on the generation prompt and a rendered image but does not yet analyze full 3D geometry, which could conceal problematic features not visible from a single viewpoint. Most importantly, the system has not yet been evaluated with real K-12 teachers or students. We plan to conduct pilot studies to assess usability, pedagogical effectiveness, and the degree to which the system effectively reduces the authoring burden for non-technical educators.